\documentclass[]{article}
\usepackage{amsmath,amssymb,amstext,amsthm,enumerate,fancyvrb,fixltx2e,hyperref,lipsum,mathtools,setspace,tikz,float,longtable,booktabs,tabularx,enumitem,xfrac,wrapfig,lipsum,rotating,fancyhdr,multirow,nameref,lineno,color,graphicx,lettrine,setspace,url,verbatim,fancyhdr,natbib,authblk}
\usepackage[justification=centering]{subcaption}

\title{Bottom-up versus top-down control and the transfer of information in complex model ecosystems}
\author[1,*]{Katharina Brinck}
\author[1,2]{Henrik Jeldtoft Jensen}
\affil[1]{Imperial College London, Department of Mathematics, Centre for Complexity Science, London SW7 2AZ}
\affil[2]{Institute of Innovative Research, Tokyo Institute of Technology, 4259, Nagatsuta-cho, Yokohama 226-8502, Japan Japan}
\date{}
\affil[*]{Corresponding author: k.brinck14@imperial.ac.uk}

\begin{document}

\maketitle


\begin{abstract}

Ecological systems are emergent features of ecological and adaptive dynamics of a community of interacting species. By natural selection through the abiotic environment and by co-adaptation within the community, species evolve, thereby giving rise to the ecological networks we regard as ecosystems. This reductionist perspective can be contrasted with the view that as species have to fit in the surrounding system, the system itself exerts selection pressure on the evolutionary pathways of the species. This interplay of bottom-up and top-down control in the development and growth of ecological systems has long been discussed, however empirical ecosystem data is scarce and a comprehensive mathematical framework is lacking.
We present a way of quantifying the relative weight of natural selection and coadaptation grounded in information theory, to assess the relative role of bottom-up and top-down control in the evolution of ecological systems, and analyse the information transfer in an individual based stochastic complex systems model, the Tangled Nature Model of evolutionary ecology.
We show that ecological communities evolve from mainly bottom-up controlled early-successional systems to more strongly top-down controlled late-successional systems, as coadaptation progresses. Species which have a high influence on selection are also generally more abundant. Hence our findings imply that ecological communities are shaped by a dialogue of bottom-up and top-down control, where the role of the systemic selection and integrity becomes more pronounced the further the ecosystem is developed.

\end{abstract}


\section*{Introduction}

Ecosystems are emergent features of ecological and adaptive dynamics of a community of interacting species \citep{Lindeman1942,Jorgensen2007}. By natural selection through the abiotic environment and by co-adaptation within the community, species evolve, thereby giving rise to the ecological networks we regard as ecosystems.

From a classical reductionist perspective, ecosystems are simply the sum of those communities, populations and organisms, which have shown to be "fittest" in the course of adaptation and selection. At the same time, ecosystems form coadapted entangled networks, where species are dependent on each other in complex manners. For a species to do well within its ecological community, it doesn't only need to win natural selection for intrinsic fitness and adaptation to the environment - it also has to fit in its ecological community; being benefited by the current system and at the same time being beneficial for the persistence of the system. Not even taking into account any metaphysical considerations, it becomes clear that the system exerts selection pressure on the evolutionary pathways of the species themselves \citep{Odum1969,Ulanowicz1986a,Jorgensen2007}.

This interplay of bottom-up and top-down control in the development and growth of ecological systems has long been discussed \citep{Ulanowicz1986a,Johnson1990,Nielsen2000c,Jorgensen2007,Brinck2017b}, however empirical ecosystem data is scarce and a comprehensive mathematical framework is lacking. We present a way of quantifying the relative weight of natural selection and coadaptation grounded in information theory, to assess the relative role of bottom-up and top-down control in the evolution of ecological systems. We use an individual based stochastic model of coevolution inspired by the Tangled Nature Model of evolutionary ecology to assess the measures over the course of ecosystem development.

The aim of this study is to assess transfer entropy as a mathematical framework for quantifying selection pressure in terms of bottom-up and top-down control in ecosystems. We expect to see differences in the relative role of bottom-up and top-down control in different successional stages. Furthermore, those species which are strongly coadapted to the current community are expected to be the ones mainly influenced by top-down control, and those species with a high contribution to the system's overall degree of coadaptation should be the main drivers of bottom-up control.

\section*{Methods}

\underline{The model}

The modelling study uses on the Tangled Nature Model of evolutionary ecology, as introduced by \citet{Christensen2002}. In previous studies, the Tangled Nature Model has shown to be able to reproduce a number of ecological patterns and distributions such as species abundance distributions, species are relationships and trophic networks \citep{Hall2002,Anderson2004,Jensen2004,Rikvold2007,Laird2008,Brinck2017b} and can hence serve as a well-tested basis for the prevailing study.

An individual is represented by a vector $S^\alpha = (S_1^\alpha, S_2^\alpha,...,S_L^\alpha)$ in the genotype space $\mathcal{S}$, where the $L$ different "genes" can take the values $\pm1$. The genotype space $\mathcal{S}$ hence represents an $L$-dimensional hypercube and encompasses all possible ways of combining the genes into a genotype sequence. There is no differentiation between genotype and phenotype. The viability of a genotype is determined by the currently perceived environment of a genotype, hence individual fitness is a function of the interactions with all other present genotypes.

The system consists of $n(S^\alpha,t)$ individuals of genotype $S^\alpha$ and $N(t)$ individuals in total. In each time step, the following dynamics are executed.

\begin{enumerate}[noitemsep]
	\item Select one individual from the pool of individuals at random and remove it from the community with probability $P_{\text{death}}$.
	\item Select one individual from the pool of individuals at random and let it reproduce with probability $P_\text{off}$. Reproduction happens asexually and the parent individual is replaced by two offspring being exact copies of the parent.
	\item Each of the both offspring undergoes mutation in each of its genes with probability $P_\text{mut}$. Mutations means that the gene value is changed from $-1$ to $1$ or vice versa.
\end{enumerate}

$P_\text{mut}$ and $P_\text{death}$ are constant and equal for all species, while $P_\text{off}$ is determined by the degree of coadaptation $\tilde{J}(S^\alpha,t)$ with the other currently present individuals,

\begin{equation}\label{equ:cs:nw:TE:Jtilde}
\tilde{J}(S^\alpha,t)  \coloneqq \frac{\sum_{S \in \mathcal{S}}J(S^\alpha,S)n(S,t)}{N(t)},
\end{equation}

where $J$ is a matrix of dimension $(2^L\times2^L)$ and stores the interaction effects for each pair of genotypes. An interaction link $J(S^\alpha,S^\beta)$ exists with probability $\theta_\text{int}$. Self interaction is zero ($J(S^\alpha,S^\alpha) = 0$), which corresponds to equal intraspecific competition across species. The non-zero entries of $J$ are for numerical convenience the product of two uniformly distributed random numbers between $-1$ and $1$ and independent for all $J(S^\alpha,S^\beta)$ (and $J(S^\beta,S^\alpha)$). $\tilde{J}(S^\alpha,t)$ can be understood as the average interaction effect of all individuals $S$ in the genotype space $\mathcal{S}$ on genotype $S^\alpha$.

Asexual reproduction occurs with probability

\begin{equation} \label{equ:cs:nw:TE:Poff}
P_\text{off}(S^\alpha,t) = \frac{\exp(\Xi(S^\alpha,t))}{1+\exp(\Xi(S^\alpha,t))} \in (0,1),
\end{equation}

where the weight function $\Xi(S^\alpha,t)$ is defined by

\begin{equation}
\Xi(S^\alpha,t) = w \cdot \tilde{J}(S^\alpha,t) - \frac{N(t)}{R}.
\end{equation}

$w$ scales the relative importance of the interaction effects for the offspring probability of a species and $R$ stands for the quality of the habitat and controls the carrying capacity of the community.

An initial population of size $N_\text{init}$ is randomly distributed over the genotype space; the initial configuration does not qualitatively influence the long-term dynamics. A generation consists of $N(t)/p_\text{death}$ time steps, which corresponds to the average time taken to kill all living individuals. The model is run for a total of $NG$ generations. Evolutionary dynamics acting on the individual genotypes give rise to species, which, within a certain regime of parameters, form long-term persisting quasi-stable mutually interacting communities (quasi-Evolutionary Stable Strategies or qESS), interrupted by brief periods of hectic reorganisation and transition to a new qESS.

To study the difference between early and late-successional communities, three different scenarios are compared. The first scenario represents the early stages of succession, an only recently disturbed system, which has to rearrange and no stable communities have formed yet. In the intermediate scenario, quasi-stable communities emerge from the interactions (see Results). In the third scenario, a late-successional enduring community is modelled. The parameter values for the scenarios are specified in Table \ref{tab:cs:nw:TE:par}.

\begin{table}[ht]\footnotesize
	\centering
	\begin{tabularx}{0.8\textwidth}{p{0.25\linewidth} X X X} \toprule
		Successional stage: & early & medium & late\\ \midrule
		$L$ & 10 & 10 & 10\\
		$\theta_\text{int}$ & 0.25 & 0.25 & 0.25\\
		$N_\text{init}$ & 100 & 100 & 100\\
		$w$ & 33 & 33 & 33\\
		$P_\text{death}$ & 0.2 & 0.2 & 0.2\\
		$P_\text{mut}$ & 0.1 & 0.01 & 0.001\\
		$R$ & 143 & 143 & 143\\ \bottomrule	
	\end{tabularx}
	\caption[Choice of model parameters for studying selection pressure]{Choice of parameters for the three scenarios. ($L$ = genome length, $\theta_\text{int}$ = probability of interaction, $N_\text{init}$ = initial number of individuals, $w$ = strength of interaction effect, $P_\text{death}$ = probability of death, $P_\text{mut}$ = gene-wise mutation probability, $R$ = carrying capacity parameter)}
	\label{tab:cs:nw:TE:par}
\end{table}

\underline{Quantifying information transfer}

Transfer entropy quantifies the information transfer between two stochastic processes and is defined as

\begin{equation}\label{equ:cs:nw:TE:TE}
TE_{X \rightarrow Y} \coloneqq \sum P(Y_{n+1},Y_n,X_n) \ln \frac{P(Y_{n+1}|Y_n,X_n)}{P(Y_{n+1}|Y_n)}.
\end{equation},

where $X$ and $Y$ are Markov processes of order 1.

To quantify the information transfer from the micro- to the macrolevel and vice versa, we need to pick two time series, representing micro- and macrodynamics. The concrete choice of time series is somewhat arbitrary, which is why we use the most straight forward choice, namely the number of individuals in a species respectively the system:

\begin{equation}
\begin{split}
m_{S^\alpha}(t) &= n(S^\alpha,t)\\
M_{S^\alpha}(t) &= N(t) - n(S^\alpha,t).
\end{split}
\end{equation}

The reason we chose to look at the difference between the total number of individuals in the system and the number of individuals within a certain species as macroscopic time series $M(t)$ is the exclusion of autocorrelation as far as possible. We use these two time series to calculate the transfer entropy from the micro- to the macrolevel $TE_{m \rightarrow M}$ and vice versa $TE_{M \rightarrow m}$. The state space is reconstructed by coarse graining of the measurements in 15 equidistant histogram bins.

\underline{Species traits}

To assess, which species are the main drivers of bottom-up control and which are those, which are most influenced by top-down control, a couple of species traits are regarded.

The average current degree of coadaptation of a species to the community is the time-average of $\tilde{J}(S^\alpha,t)$

\begin{equation}\label{equ:cs:nw:TE:Jinave}
\overline{J_\text{in}(S^\alpha)}  \coloneqq \frac{1}{NG} \cdot \sum_{t=1}^{NG}\tilde{J}(S^\alpha,t),
\end{equation}

as $\tilde{J}(S^\alpha,t)$ denotes the effect of all individuals in the community on an individual of species $S^\alpha$ (compare equ. \ref{equ:cs:nw:TE:Jtilde}).

The average current degree of contribution to the coadaptation of the community is defined as the sum of all effects an individual of species $S^\alpha$ has on the individuals of the community

\begin{equation}\label{equ:cs:nw:TE:Joutave}
\begin{split}
\overline{J_\text{out}(S^\alpha)}  &\coloneqq \frac{1}{NG} \cdot \sum_{t=1}^{NG}J_\text{out}(S^\alpha,t)\\
J_\text{out}(S^\alpha,t) &\coloneqq \frac{\sum_{S \in \mathcal{S}}J(S,S^\alpha)n(S,t)}{N(t)}.
\end{split}
\end{equation}

Equ. \ref{equ:cs:nw:TE:Jinave} and \ref{equ:cs:nw:TE:Joutave} denote the effect of the system on the individual and vice versa. One might think of a sunflower in a typical ecosystem. Equ. \ref{equ:cs:nw:TE:Jinave} then summarises the effect of the pollinating bees, the snails trying to eat the stem of the flower, the surrounding threes that might compete for light etc. In contrast, equ. \ref{equ:cs:nw:TE:Joutave} measures the effect the sunflower has on other species such as providing honey to the bees, seed to the squirrels or shade to the slugs.

As a third possible explanatory for driving bottom-up or top-down control, we look at the average number of individuals in a species

\begin{equation}\label{equ:cs:nw:TE:nave}
\overline{n(S^\alpha)}  \coloneqq \frac{1}{NG} \cdot \sum_{t=1}^{NG}n(S^\alpha,t).
\end{equation}

\underline{Statistical methods}

To assess, if the direction of information transfer differs between the successional stages, the statistical method of effect sizes is applied. It describes the strength or relative magnitude of a phenomenon and quantifies, whether an effect "matters". Contrary to statistical significance, where with large sample sizes, very small effects can stand out as significant, the method of effect sizes reduces false positives. Cohen's $d$ for one sample is used to assess whether the transfer entropies are effectively larger than zero, and effect sizes for multiple groups are analysed to assess the effect of the successional stage on the information transfer \citep{Cohen1988,Lenhard2016}.

To analyse which species are the main drivers of bottom-up selection and are most affected by top-down selection, a linear model based on species abundance $\overline{n(S^\alpha)}$, species coadaptation $\overline{J_\text{in}(S^\alpha)}$ and species contribution to the system's coadaptation $\overline{J_\text{out}(S^\alpha)}$ is created \citep{Crawley2005}.

\section*{Results}

\underline{Dynamics}

\begin{figure}[ht]
	\centering
	\captionsetup[subfigure]{justification=centering}
	\begin{subfigure}{\linewidth}
		\centering
		\includegraphics[width=	\linewidth]{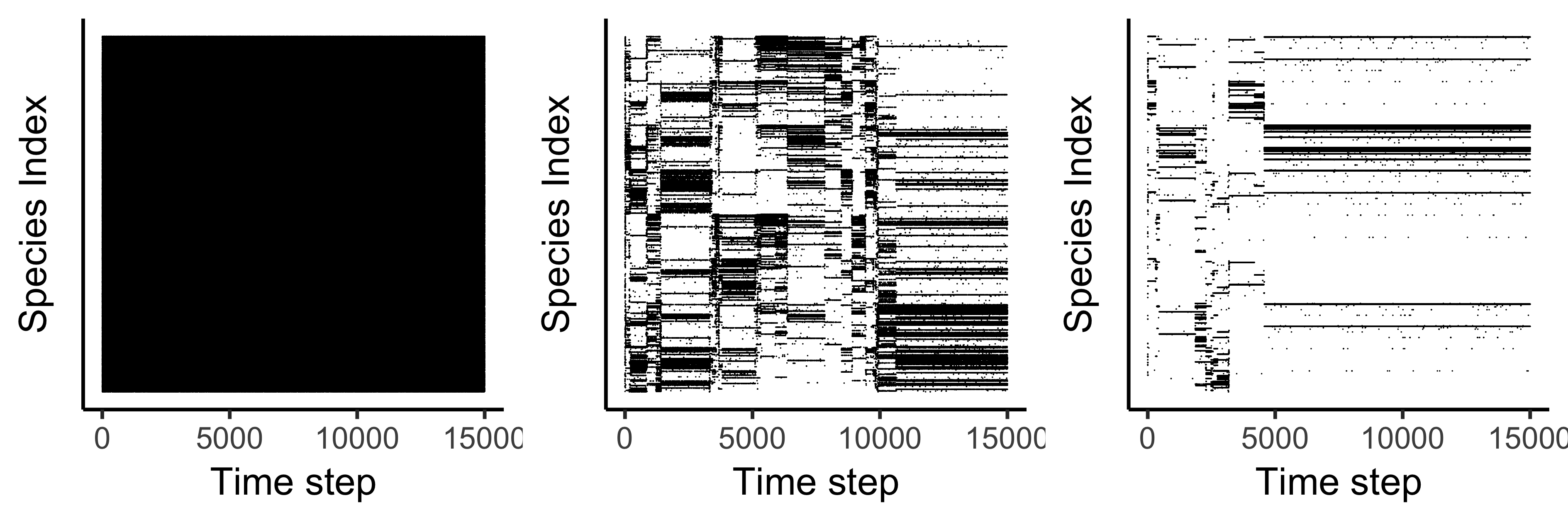}	
	\end{subfigure}
	\begin{subfigure}{0.333\linewidth}
		\centering
		\caption[]{Early successional stage}
		\label{fig:cs:nw:TE:scenarios:hectic}
	\end{subfigure}%
	\begin{subfigure}{0.333\linewidth}
		\centering
		\caption[]{Medium successional stage}
		\label{fig:cs:nw:TE:scenarios:qEES}
	\end{subfigure}%
	\begin{subfigure}{0.333\linewidth}
		\centering
		\caption[]{Late successional stage}
		\label{fig:cs:nw:TE:scenarios:stable}
	\end{subfigure}
	\caption[Three scenarios for different successional stages]{Three model scenarios, representing early (a), medium (b) and late successional (c) stages. In the early successional stages of the system, no stable communities could yet have been formed, which is why many different species enter and exit the community. The more coadapted the system becomes, the longer quasi-stable communities persist. The scenarios are parametrised according to Table \ref{tab:cs:nw:TE:par}.}
	\label{fig:cs:nw:TE:scenarios}
\end{figure}

The model behaviour for the three different scenarios is visualised in Fig. \ref{fig:cs:nw:TE:scenarios}. In the early successional stages of the system, no stable communities could yet have been formed, which is why many different species enter and exit the community, giving rise to hectic dynamics (Fig. \ref{fig:cs:nw:TE:scenarios:hectic}). The more coadapted the system becomes, the longer quasi-stable communities persist (Fig. \ref{fig:cs:nw:TE:scenarios:qEES}), until a fairly stable and persistent late successional community is formed (Fig. \ref{fig:cs:nw:TE:scenarios:stable}).

\underline{The behaviour of transfer entropy in the different successional stages}

\begin{figure}[H]
	\centering
	\captionsetup[subfigure]{justification=centering}
	\begin{subfigure}{\linewidth}
		\centering
		\includegraphics[width=	\linewidth]{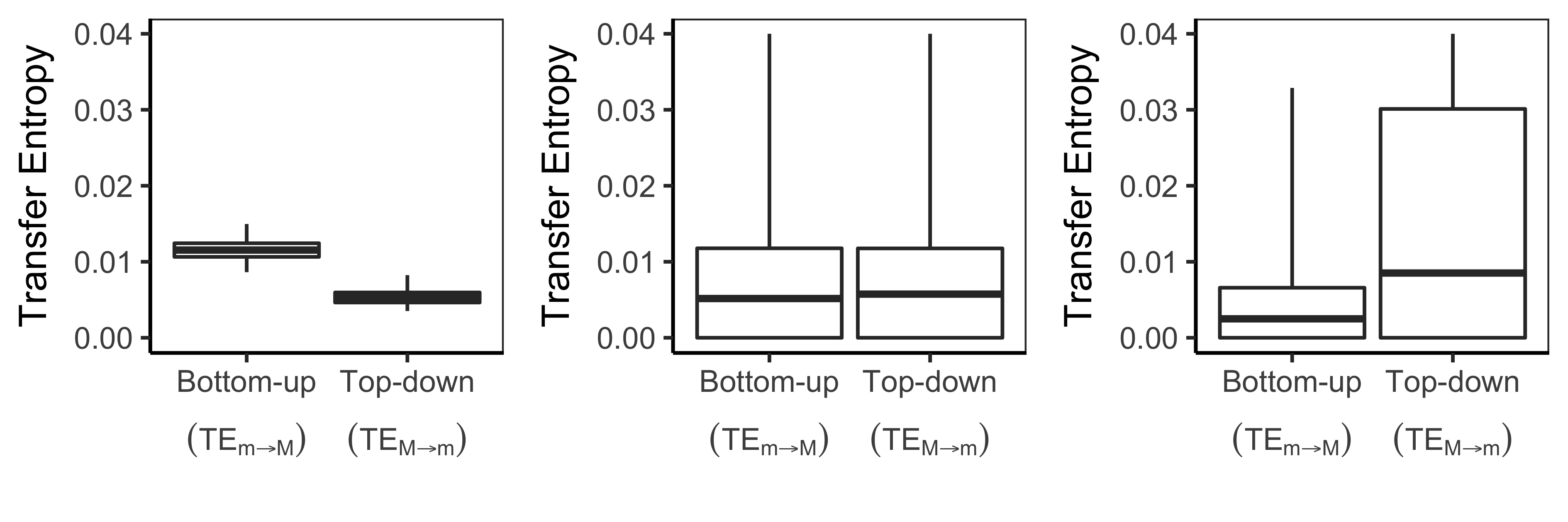}	
	\end{subfigure}
	\begin{subfigure}{0.333\linewidth}
		\centering
		\caption[]{Early successional stage}
	\end{subfigure}%
	\begin{subfigure}{0.333\linewidth}
		\centering
		\caption[]{Medium successional stage}
	\end{subfigure}%
	\begin{subfigure}{0.333\linewidth}
		\centering
		\caption[]{Late successional stage}
	\end{subfigure}
	\caption[Selection transfer in the different successional stages]{Distribution of selection transfer from the micro- to the macrolevel (bottom-up control, $TE_{m \rightarrow M}$) and vice versa (top-down control, $TE_{M \rightarrow m}$) over the set of species in the community, measured in terms of transfer entropy for the different successional stages. The boxes show the mean (solid line in the middle of the box) $\pm$ the standard deviation (boundaries of the box) as well as the minimum and maximum value (end of the solid dash). Note that the y-axis is capped; the maximum transfer entropy in the medium successional stage is 0.07, and in the late successional stage it is 0.11.}
	\label{fig:cs:nw:TE:boxes}
\end{figure}

The information transfer between the microscopic and the macroscopic time series are small, but detectably larger than zero, for all successional stages and both directions (Table \ref{tab:cs:nw:TE:effects}b). Bottom-up control is highest in the early successional stage and decreases over the course of ecosystem development (Fig. \ref{fig:cs:nw:TE:boxes}a to c). The opposite is true for top-down control: here information transfer increases on average, the more stable the system becomes. Overall, the successional stage has a strong effect on bottom-up and a small effect on top-down information transfer (Table \ref{tab:cs:nw:TE:effects}c). These results support our hypothesis and reveal a trend towards weaker bottom-up and stronger top-down control over the course of ecological succession. The differences in transfer entropy comparing bottom-up and top-down information transfer as well as the differences comparing the successional stages are both significant (ANOVA, $p<2e^{-16}$, $n = 1024 \times 6$).

\begin{table}[H]\footnotesize
	\centering
	\begin{tabularx}{\textwidth}{X X X X} \toprule
		& \boldmath$TE_{m \rightarrow M}$ & \boldmath$TE_{M \rightarrow m}$ & \boldmath$\Delta$\\ \midrule
		\multicolumn{4}{l}{\textbf{(a) Means \boldmath$\pm$ Standard Deviation}}\\ \midrule 
		Early & 0.0115  $\pm$ 0.0009 & 0.0053  $\pm$ 0.0007 & $\phantom{\text{-}}$0.0062  $\pm$ 0.0010\\
		Medium & 0.0052  $\pm$ 0.0066 & 0.0057  $\pm$ 0.0060 & -0.0006  $\pm$ 0.0069\\
		Late & 0.0025  $\pm$ 0.0041 & 0.0085  $\pm$ 0.0216 & -0.0060  $\pm$ 0.0200 \\ \midrule
		\multicolumn{3}{l}{\textbf{(b) Effect size compared to \boldmath$TE = 0$}} & \textbf{Effect size of $\Delta$}\\ \midrule
		Early & 12.886 (***) & 8.009 (***) & $\phantom{\text{-}}$7.690 (***)\\
		Medium & $\phantom{1}$0.781 (**) & 0.954 (***) & -0.079\\
		Late & $\phantom{1}$0.609 (**) & 0.394 (*) & -0.386 (*)\\ \midrule
		\multicolumn{4}{l}{\textbf{(c) Effect size of the successional stage on the \boldmath$TE$}}\\ \midrule 
		& $\phantom{1}1.536$ (***) & $0.313$ (*) & \\ \bottomrule	
	\end{tabularx}
	\caption[Effect sizes of selection transfer entropies]{Mean and standard deviation of the selection transfer entropies in the different scenarios as well as effect sizes. $\Delta \coloneqq TE_{m \rightarrow M}-TE_{M \rightarrow m}$, (*): small effect, (**): intermediate effect, (***): large effect \citep[according to][]{Cohen1988}}
	\label{tab:cs:nw:TE:effects}
\end{table}

\underline{Which species drive bottom-up respectively top-down control?}

Fig. \ref{fig:cs:nw:TE:boxes} also gives an indication about the spread of the influences different species have on the bottom-up and top-down information transfer. While in the early successional stage, most species are lowly populated, barely coadapted and hence have similar influence on the information transfer, in the medium and especially late successional stage, a few species lie far higher than the median and have a comparably large effect on bottom-up control respectively are strongly affected by top-down control.

\begin{figure}[H]
	\centering
	\captionsetup[subfigure]{justification=centering}
	\begin{subfigure}{\linewidth}
		\centering
		\includegraphics[width=	\linewidth]{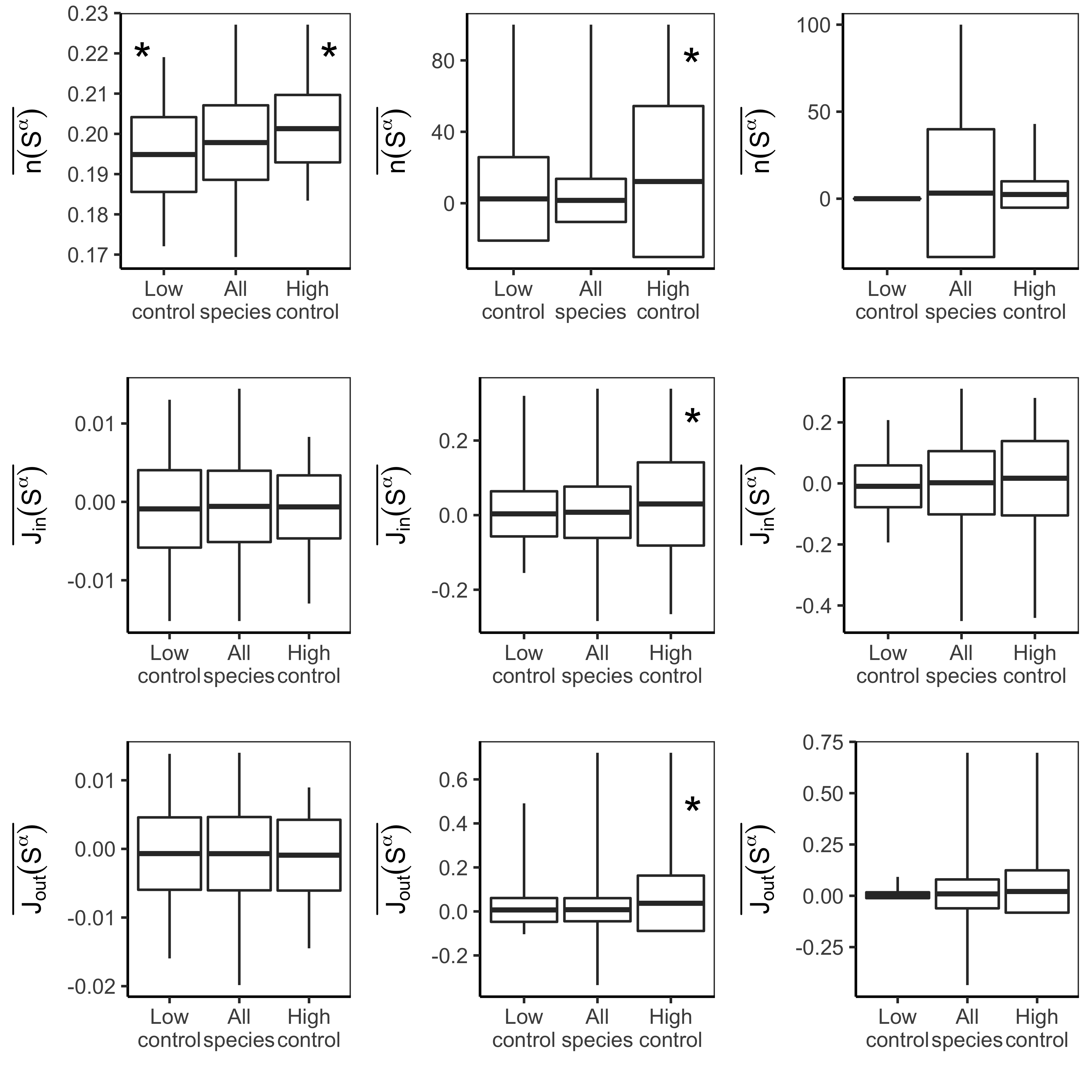}
	\end{subfigure}
	\begin{subfigure}{0.33\linewidth}
		\centering
		\caption[]{Early successional stage}
	\end{subfigure}%
	\begin{subfigure}{0.33\linewidth}
		\centering
		\caption[]{Medium successional stage}
	\end{subfigure}%
	\begin{subfigure}{0.33\linewidth}
		\centering
		\caption[]{Late successional stage}
	\end{subfigure}
	\caption[Comparison of species with high and low bottom-up and top-down control]{Trait distribution of species with low and high bottom-up and top-down control compared to the overall set of species in the community. Regarded traits are abundance $\overline{n(S^\alpha)}$, species coadaptation $\overline{J_\text{in}(S^\alpha)}$ and species contribution to the system's coadaptation $\overline{J_\text{out}(S^\alpha)}$. The boxes show the mean (solid line in the middle of the box) $\pm$ the standard deviation (boundaries of the box) as well as the minimum and maximum value (end of the solid dash). A species is defined as having a "high control"/"low control" effect, if its respective transfer entropy is in the upper/lower 5\% percentile of the respective transfer entropies in the population. Significant deviations from the overall species community means are marked with an asterisk.}
	\label{fig:cs:nw:TE:frequdists}
\end{figure}

Fig. \ref{fig:cs:nw:TE:frequdists} shows, how the traits of species, which have a particularly high or low bottom-up or top-down control, compare to the overall distribution of species traits in the community (compare equ. \ref{equ:cs:nw:TE:nave}, \ref{equ:cs:nw:TE:Jinave} and \ref{equ:cs:nw:TE:Joutave}). It can be seen, that the high control-species are generally more abundant (upper row), better coadapted (middle row) and have a stronger influence on community adaptation (lower row). However significant deviations from the overall species pool mean only appear in the medium successional stage, and for abundance also in the early-successional stage.

To analyse, which of the species traits introduced above best explain how much a species contributes to the transfer entropies, a linear regression model with the average number of individuals in the species, the average degree of coadaptation and the average contribution of the community's coadaptation is constructed (Table \ref{tab:cs:nw:TE:whichspecies}). Different than expected, there is no clear distinction between which species traits explain high bottom-up and which strong top-down control. In the early successional stage, the abundance and degree of coadaptation of a species determine its effect on bottom-up and top-down control. In the medium successional stage, the coadaptation effect of and on a species significantly influence top-down control, whereas in late successional stages, the same is true for bottom-up control.

\begin{table}[H]\footnotesize
	\includegraphics[width=	\linewidth]{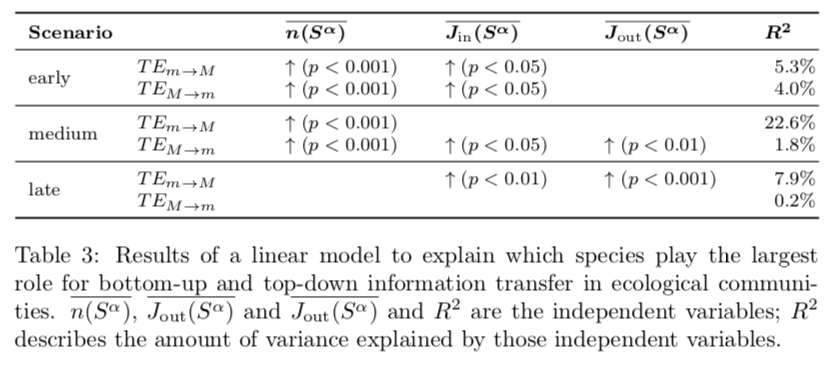}
	\label{tab:cs:nw:TE:whichspecies}
\end{table}

\section*{Conclusions}

Fitness is a multidimensional measure involving intrinsic and extrinsic parameters and can be regarded from two sides - how well does a species do in the current context, and how much does a species contribute to sustain the current environment. The information theoretic concept of transfer entropy can serve as a mathematical framework to capture these aspects of top-down and bottom-up selection in ecosystems.

We have demonstrated, that bottom-up and top-down control are of different importance over the course of ecosystem succession. While early successional stages are dominated by bottom-up control, late successional systems are more strongly influenced by top-down control (Fig. \ref{fig:cs:nw:TE:boxes}). This makes sense as in early successional communities, no coadapted subset of species has become dominant yet and most species have similar influence on the community. The community is simply comprised of those sunflowers, pollinators or slugs, which happen to do well intrinsically and are able to acquire enough resources. Over the course of succession, the ecological communities become more coadapted and being competitive isn't sufficient any more. Whether a new pollinator or species of flower is able to establish in the community depends on how well it interacts with the present community, how much positive interactions it experiences and how positively it contributes to the community.

With progressing developmental stage, the variance in contribution to information transfer across species increases (Fig. \ref{fig:cs:nw:TE:boxes}). While in the early successional stage, the number of individuals and the amount of benefit a species perceives \emph{from} the community ($J_\text{in}(S^\alpha)$) predominantly explains a species' contribution to information transfer, in medium and late successional stages also the contribution \emph{to} the community ($J_\text{out}(S^\alpha)$) becomes a significant explanatory variable for effect on information transfer.

These findings imply, that more developed ecological systems do indeed exert selection pressure on the present species - without the need for any metaphysical macroscopic optimisation criteria, group selection hypotheses or anything related, but only via information transferring feedbacks within the system.

Further research on different models and ideally also empirical data has to show if the concept of transfer entropy can serve as a solid basis for a rigid mathematical framework to quantify the roles of bottom-up and top-down control in different model, experimental and real-world ecosystems. If so, it can shed light on general selection principles and eventually help to forecast and potentially prevent periods rapid change accompanied by mass extinctions in real-world ecosystems.

\section*{Acknowledgements}
Simulations have been run on the High Performance Cluster by the Imperial College Computing Service whom we thank sincerely for providing these facilities. KB thanks Imperial College’s Department of Mathematics for funding her PhD work.

\newpage
\bibliographystyle{elsarticle-harv}


\end{document}